\documentclass[pra,twocolumn,showpacs,superscriptaddress,floatfix]{revtex4}
\usepackage[dvipdfm]{graphicx}
\usepackage{subfigure}
\usepackage{epsfig}
\usepackage{float}
\usepackage{ifpdf}
\ifpdf
	\usepackage[pdftex]{graphicx}
	\usepackage{epstopdf}
\else
	\usepackage[dvipdfm]{graphicx}
\fi
\usepackage{amsmath}	
\newcommand{\abs}[1]{\left| #1 \right|}
\newcommand{\ket}[1]{\left | #1 \right \rangle}
\def\k(#1){|#1\rangle}
\newcommand{\bra}[1]{\left \langle #1 \right |}
\newcommand{\proj}[1]{\ket{#1} \bra{#1}}

\newcommand{\beq}{\begin{equation}}
\newcommand{\eeq}{\end{equation}}
\newcommand{\beqa}{\begin{eqnarray}}
\newcommand{\eeqa}{\end{eqnarray}}
\newcommand{\beqan}{\begin{eqnarray*}}
\newcommand{\eeqan}{\end{eqnarray*}}

\newcommand{\affA}{%
\affiliation{
 National Institute of Information and Communications Technology,
 4-2-1 Nukui-kita, Koganei, Tokyo 184-8795, Japan}
     }
\newcommand{\affB}{%
\affiliation{
 Sophia University,
 7-1 Kioicho, Chiyoda-ku, Tokyo 102-8554, Japan}
}

\newcommand{\affC}{%
\affiliation{
 Raytheon BBN Technologies, 
 10 Moulton Street, Cambridge, MA 02138, USA}

}

\begin{document}

\title{\bf Quantum receivers with squeezing and photon-number-resolving 
detectors \\
for $M$-ary coherent state discrimination 
}

\date{\today}

\author{Shuro Izumi }
\affA \affB
\author{Masahiro Takeoka}%
\affA
\affC
\author{Kazuhiro Ema}%
\affB
\author{Masahide Sasaki}%
\affA


\begin{abstract}
We propose quantum receivers with optical squeezing and photon-number-resolving detector (PNRD) 
for the near-optimal discrimination of quaternary phase-shift-keyed coherent state signals. 
The basic scheme is similar to the previous proposals 
(e.g. Phys.\ Rev.\ A {\bf 86}, 042328 (2012)) 
in which displacement operations, on-off detectors, 
and electrical feedforward operations were used. 
Here we study two types of receivers where one installs
optical squeezings and the other uses PNRDs instead of on-off detectors. 
We show that both receivers can attain lower error rates than that by
the previous scheme.  In particular, we show the PNRD based receiver has 
a significant gain when the ratio between the mean photon number of 
the signal and the number of the feedforward steps is relatively high, 
in other words, the probability of detecting two or more photons 
at each detector is not negligible.
Moreover, we show that the PNRD based receiver can suppress the errors 
due to dark counts, which is not possible by the on-off detector based 
receiver with a small number of feedforwards. 
\end{abstract}

\maketitle

\section{Introduction}
Coherent states are known as the best signal carriers 
in optical communication owing to their loss-tolerant property. 
Coherent states propagating through a lossy channel remain
in a pure coherent state with decreased amplitude while 
more exotic states such as photon number states easily lose their purity 
in a lossy channel. 
In fact, the ultimate channel capacity in a lossy bosonic channel 
can be attained by using coherent state carriers, appropriate classical 
encoding and a quantum mechanically 
optimal decoding (measurement)
 over a sequence of coherent states \cite{Giovannetti04}.

Implementation of such kind of quantum collective decoding 
is still challenging.
Currently, attentions are paid to 
implementing quantum receivers for detecting each coherent pulse separately
at a smaller error rate than the conventional limit({\it standard quantum limit} : SQL)
which is attained by homodyne/heterodyne receivers.

The quantum mechanical bound of the minimum error rate is called the Helstrom 
bound and is known to be significantly lower than the SQL. 
Physical implementations of the optimal quantum receiver to achieve
the Helstrom bound were studied theoretically in the 1970's by Kennedy and Dolinar \cite{Kennedy73,Dolinar73,Helstrom_book76_QDET} 
and recently further investigated from a more practical point of view
\cite{Sasaki96_Unitary_Control,Geremia04,Takeoka05,TakeokaSasakiLutkenhaus2006_PRL_BinaryProjMmt,TakeokaSasaki2008_DisplacementRec_GaussianLimit,ADP11}.
Experimentally the super-SQL performances even without compensating for imperfections were demonstrated for both 
on-off-keying (OOK) \cite{CookMartinGeremia2007_Nature,Tsujino2010_OX_OnOff} 
and binary phase-shift-keying (BPSK) \cite{Wittmann2008_PRL_BPSK,Tsujino2011_Q_Receiver_BPSK}. 

More recently, attention has also been paid to the $M$-ary signal with $M>2$ 
where one can encode the message in pulses more densely \cite{Bondurant93,Becerra_NIST_2011_MPSK_emulation_experiment,izumi2012,Becerra13,Mueller2012_NJP,Osaki_Ban_Hirota,Nair_and_Yen}. 
Bondurant \cite{Bondurant93} extended Dolinar's optimal binary signal 
discrimination scheme and proposed near-optimal receivers for 
quaternary phase-shift-keying (QPSK), which consists of a local oscillator, 
an on-off detector (which distinguish only zero or non-zero photons), 
and an infinitely fast feedback operation. 
Later, more practical schemes assuming the finite number of feedforward steps 
have been studied \cite{Becerra_NIST_2011_MPSK_emulation_experiment,izumi2012}
and very recently the super-SQL performance was experimentally demonstrated 
\cite{Becerra13}. 
As another approach, the QPSK discrimination by 
a hybrid receiver of homodyne and on-off detections was also proposed 
\cite{Mueller2012_NJP}. 
In addition, the near-optimal discrimination was also demonstrated 
for the pulse-position coding \cite{Guha2011_JMO,Chen2012_NatPh}.
In all of these schemes, 
the near-optimal performance is achieved by inducing 
effective optical nonlinearities 
via an on-off detection and the ultrafast electrical feedback 
(or feedforward) operation.

In this paper, we theoretically show that additional 
optical nonlinear processes, squeezing and photon-number-resolving detector 
(PNRD), are also useful for the QPSK coherent state 
discrimination. 
For the binary case, it was shown that the squeezing can slightly 
improve the error rate performance 
\cite{TakeokaSasaki2008_DisplacementRec_GaussianLimit}. 
We show that the similar effect can be observed by installing 
squeezers into the QPSK receiver scheme we proposed previously 
\cite{izumi2012}. 

PNRD is also known as an attractive device to induce an effective optical 
nonlinearity in optical quantum information processing \cite{Bartlett02}. 
For the coherent state discrimination task, 
PNRD has been applied to implement a generalized (non-projective) 
measurement of discriminating binary states 
with an inconclusive result \cite{Wittmann10-1,Wittmann10-2}. 
Furthermore, a benefit of employing PNRD for $M$-ary coherent state discrimination was also implied in \cite{Becerra_NIST_2011_MPSK_emulation_experiment}.
Here we apply PNRD into the QPSK receiver based on \cite{izumi2012} 
and show that it can achieve the near-optimal error performance
even if the number of feedback steps is relatively small. 
In other words, PNRDs can decrease the number of feedforward steps 
to attain the same performance. 
In addition, we show that this scheme is highly robust against 
the dark counts.

In Sec.~\ref{Sect:2}, we propose and analyze the QPSK receiver 
with the squeezing operation. 
The receiver with PNRD is proposed and numerically simulated 
in Sec.~\ref{Sect:3}. 
Section \ref{Sect:4} concludes the paper. 

\section{Displacement receiver with squeezing operation}\label{Sect:2}

In this section, we apply the squeezing operations into 
the displacement receiver proposed in \cite{izumi2012}. 
The signals to be discriminated are $M$-PSK coherent states defined as 
\beq 
\ket{\alpha_m}=\ket{\alpha\, u^m},\quad u=e^{2\pi i / M}\;,
\label{eq:1}
\eeq
where $m=0,1,\ldots,M-1$ and 
$\alpha$ is chosen to be real without loss of generality. 
Throughout this paper, we fix $M=4$ and assume 
equal {\it a priori} probabilities, i.e. $p_m=1/M$ for all $m$. 

\begin{figure}[h]
\begin{center}
\includegraphics[width=0.9\linewidth]
{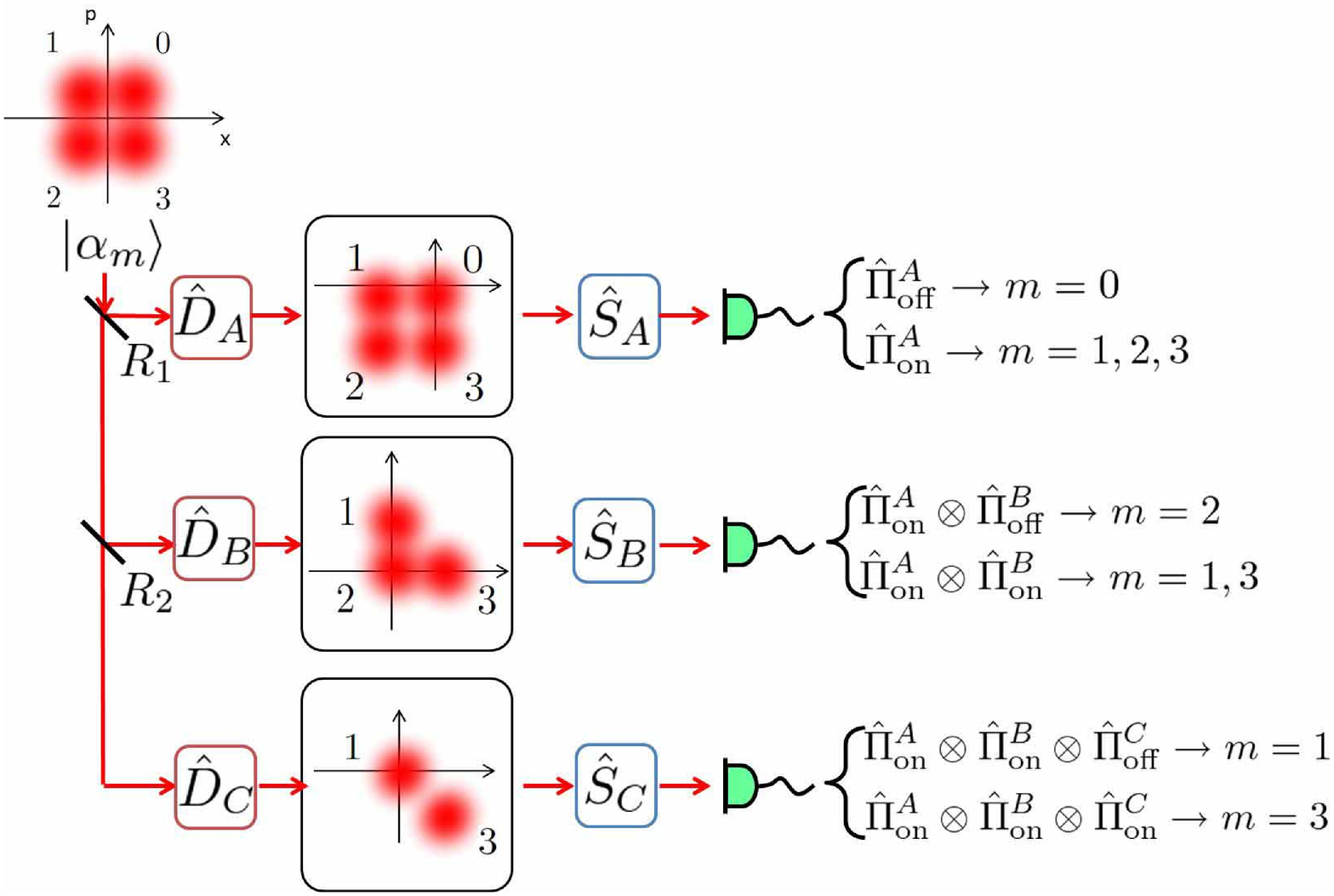}
\caption{(Color online) Displacement receiver with three-port detection structure
without feedforward operations.
The squeezing operations are applied to the displaced signals.
\label{Scheme_4PSK_3port}
}
\end{center}
\end{figure}

A schematic of the receiver is depicted in Fig.~\ref{Scheme_4PSK_3port}. 
It consists of beam splitters, displacement operations, squeezing, 
and on-off detectors. 
The QPSK signal is split into three ports via two beam splitters having reflectance $R_{1}$ and $R_{2}$, respectively.
At each port, the displacement operation is applied. 
The operator is given by 
$ \hat{D}(\gamma)=\exp[(\gamma \hat{a}^{\dagger} - \gamma^{\ast} \hat{a})] $ 
which shifts the amplitude of coherent state as 
$\hat{D}(\gamma)|\beta\rangle = |\beta+\gamma\rangle$. 
It is widely known that $\hat{D}(\gamma)$ is 
realized by a beam splitter with transmittance $\tau \approx 1$ and relatively strong local oscillator $\ket{\gamma/\sqrt{\tau}}$. 
In our scheme it 
is applied such that one of the four symbols is displaced to be close to 
the vacuum state (signal nulling). 
The previous works \cite{TakeokaSasaki2008_DisplacementRec_GaussianLimit,Wittmann2008_PRL_BPSK,Tsujino2010_OX_OnOff,Tsujino2011_Q_Receiver_BPSK,izumi2012,Mueller2012_NJP,Guha2011_JMO,Chen2012_NatPh} showed 
that the optimal displacement minimizing the average error 
is slightly different from the exact nulling, 
which is taken into account in here as well.
The target signals to be nulled at port $A$, $B$, and $C$ are $m = 0$, 2, 1.

The displaced signal at each port is then squeezed by 
the squeezing operation $\hat{S}(\xi)=\exp[(\xi^{\ast} \hat{a}^2-\xi \hat{a}^{\dagger 2})/2]$, where $\xi=r \mathrm{e}^{i \phi}$ is the complex squeezing parameter, and then detected by an on-off detector which distinguish if the signal 
is the nulled one or not. 
The on-off detection process at three ports are 
described by an appropriate set of three-mode 
measurement operators $\{ \hat{\Pi}_i \}$ ($i=0,1,2,3$) and 
the correct detection probabilities are then given by 
\begin{equation}
P(i\vert i)=\bra{\Psi_i}\hat\Pi_i\ket{\Psi_i}. 
\label{eq:success_prob}
\end{equation}
From them, we obtain the average error probability as 
\begin{eqnarray}
P_e=1-\frac{1}{4}\sum_{i=0}^{3} P(i|i)\;.
\label{eq:7}
\end{eqnarray}
The parameters of the beam splitters, the displacements, and 
the squeezings are numerically optimized to minimize $P_e$. 
Details of the derivation of Eq.~(\ref{eq:7}) and the optimized parameters 
are described in Appendix A.

In Fig.~\ref{DR_Sq_DE_1_DC_0_n=6}, 
we plot the average error rate of the proposed receiver 
which is compared with that of the receiver without squeezing 
\cite{izumi2012}, heterodyne receiver, and the Helstrom bound. 

\begin{figure}[t]
\begin{center}
\includegraphics[width=0.9\linewidth]
{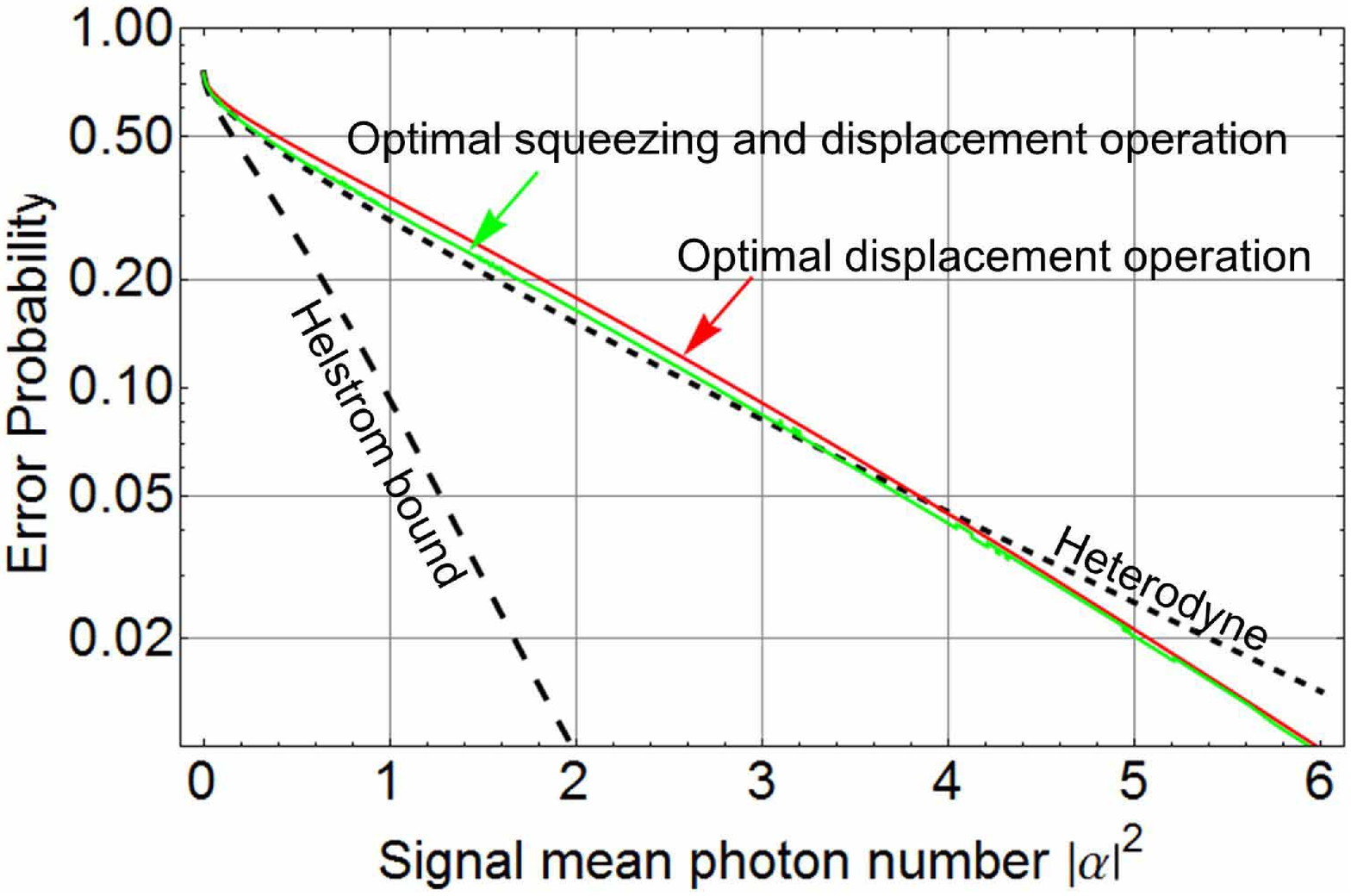}
\caption{(Color online) Average error rates for QPSK signal discrimination without applying the feedforward.  The optimized displacements and squeezings receiver (green line) and the optimized displacements receiver (read line).
$\eta=1$ and $\nu=0$.
Throughout this paper, the Helstrom bound and the heterodyne limit are represented by the black dashed and dotted line respectively.
}
\label{DR_Sq_DE_1_DC_0_n=6}
\end{center}
\end{figure}
We observe an improvement of the performance by introducing squeezing 
operations in the small photon number region $\abs{\alpha}^2 \le 4$. 
However, the improvement is extremely small and thus it is expected 
that the gain would disappear when the system imperfections 
are taken into account. 
To observe a significant gain in this approach, 
one may need higher order nonlinear optical processes. 

\section{Displacement receiver with photon number resolving detectors}\label{Sect:3}

According to Ref.~\cite{izumi2012} the error rate for QPSK can be drastically improved by employing and repeating the feedforward that is based on the binary information from the on-off detector, which allows us to set the value of the displacement at the $j$th branch $\hat D_j(\cdot)$ depending on the outcome from the $(j-1)$th branch.
Additionally, the error rate performance of the feedforward receiver can be further increased by refining 
the feedforward rule by adopting the maximization of
{\it a posteriori} probabilities.
However, the {\it a posteriori} probabilities can be more precisely estimated by replacing the conventional on-off detectors with the PNRDs,
because the amplitudes of signals are different from each other.

 In this section, we study the displacement receiver allowing the use of
PNRDs instead of on-off detectors and the feedforwards based on the Bayesian updating. 
The measurement operator of the PNRD for the $n$-photon detection is described by \cite{BarnettPhillipsPegg1998},
\begin{equation}
\hat{\Pi}_n =\mathrm{e}^{-\nu}\sum_{l=0}^{n}\sum_{k=n-l}^{\infty}\frac{\nu^{l}}{l!} C^{k}_{n-l} \eta^{n-l} (1-\eta)^{k-(n-l)}\proj{k},
\label{eq:8}
\end{equation}
where $C^{k}_{n-l}$ is the binomial coefficient.
The probability of detecting $n$ photons for the coherent state input $\ket{\beta}$ is thus given by 
\begin{eqnarray}
P(n|\beta )&=& \bra{\beta} \hat{\Pi}_n \ket{\beta}
\nonumber
\\
&=&
\mathrm{e}^{-\nu-\eta\abs{\beta}^2} \frac{(\nu+\eta\abs{\beta}^2)^n}{n!}.
\label{eq:9}
\end{eqnarray}

\begin{figure}[b]
\begin{center}
\includegraphics[width=0.9\linewidth]
{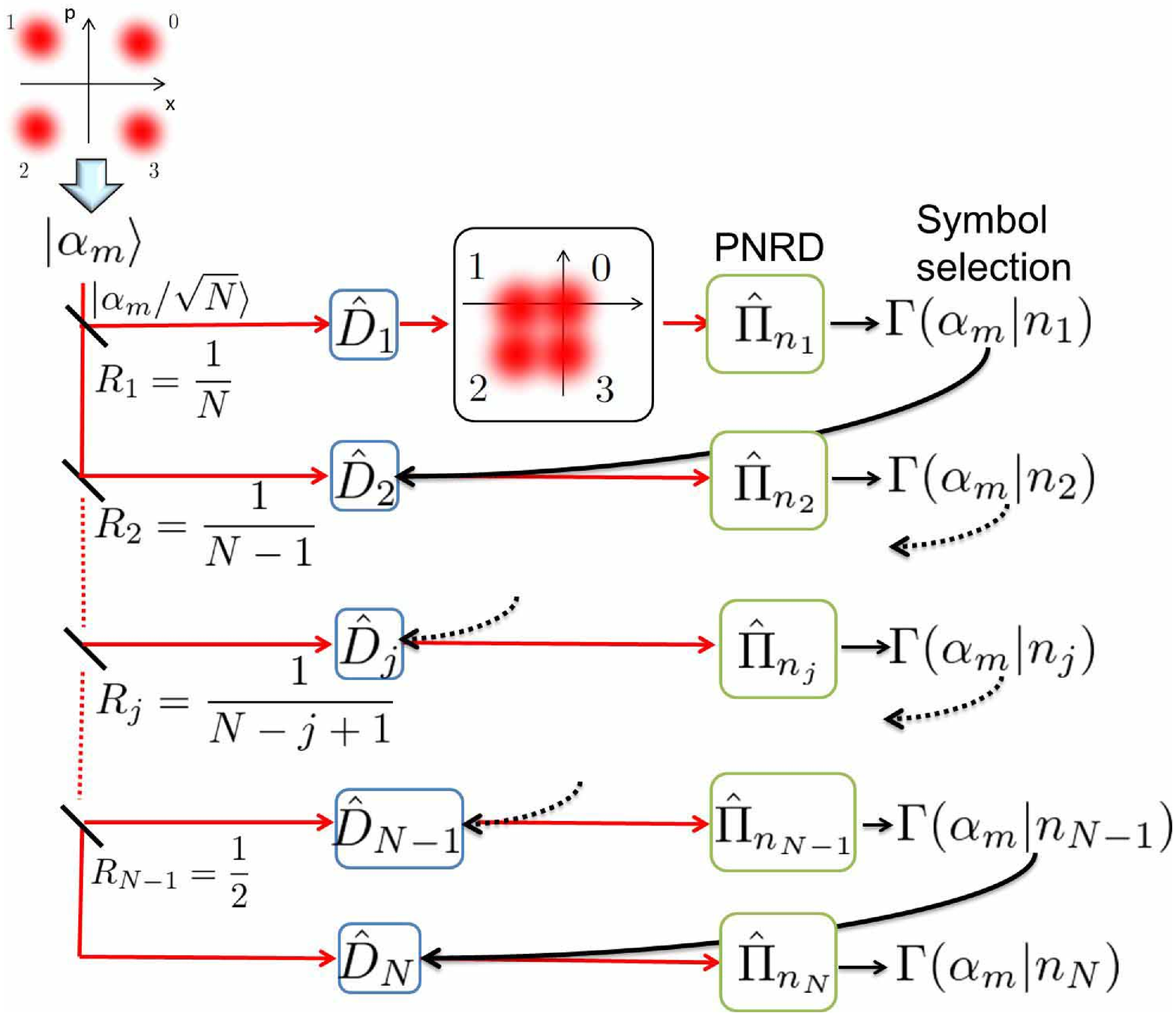}
\caption{(Color online) The displacement receiver consisting of $N$-step feedforward operations.
The nulling symbol $m_j$ on the $j$th stage is determined by the {\it a posteriori} probabilities on the $(j-1)$th stage. 
}
\label{detection_scheme_N_feed}
\end{center}
\end{figure}

The schematic of the receiver is shown in Fig.~\ref{detection_scheme_N_feed}, 
which is similar to those described in \cite{izumi2012,Becerra_NIST_2011_MPSK_emulation_experiment} except that the on-off detectors are replaced by the PNRDs. 
The input signal is equally split into $N$ ports via $N-1$ beam splitters. 
We denote the nulling symbol on the $j$th stage as $m_{j}$.
The nulling symbol at the first port is set to be $m_{1}=0$ while the symbols $m_{j}$ ($j \ge 2$) are chosen to have the maximum 
{\it a posteriori} probability. 
The {\it a posteriori} probability for after detecting $n_j$ photons 
at the $j$th stage is given by 

\begin{eqnarray}
\Gamma ( \alpha_{m}|n_{j})
&=&
\frac{\Gamma ( \alpha_{m}|n_{j-1})P(n_{j}|(\alpha_{m}-\alpha_{m_{j}})/\sqrt{N})}{\sum_{l=0}^{M}\Gamma ( \alpha_{l}|n_{j-1})P(n_{j}|(\alpha_{l}-\alpha_{m_{j}})/\sqrt{N})}
\nonumber
\\
&=&
\frac{p_{m}\Pi_{h=1}^{j} P(n_{h}|(\alpha_{m}-\alpha_{m_{h}})/\sqrt{N})}{\sum_{l=0}^{M}p_{l}\Pi_{h=1}^{j} P(n_{h}|(\alpha_{l}-\alpha_{m_{h}})/\sqrt{N})},
\label{eq:10}
\end{eqnarray}
where $p_{m}$ denotes the {\it a prior} probability.

We first derived an analytical expression of the average error rate assuming $\nu=0$. 
In this case, 
once photons are detected from the signal in which $m_j$ is nulled, 
we have $\Gamma(\alpha_{m_{j}}|n_{j})=0$ and thus 
Eq.~(\ref{eq:10}) is drastically simplified.

\begin{figure}[h]
\begin{center}
\includegraphics[width=0.9\linewidth]
{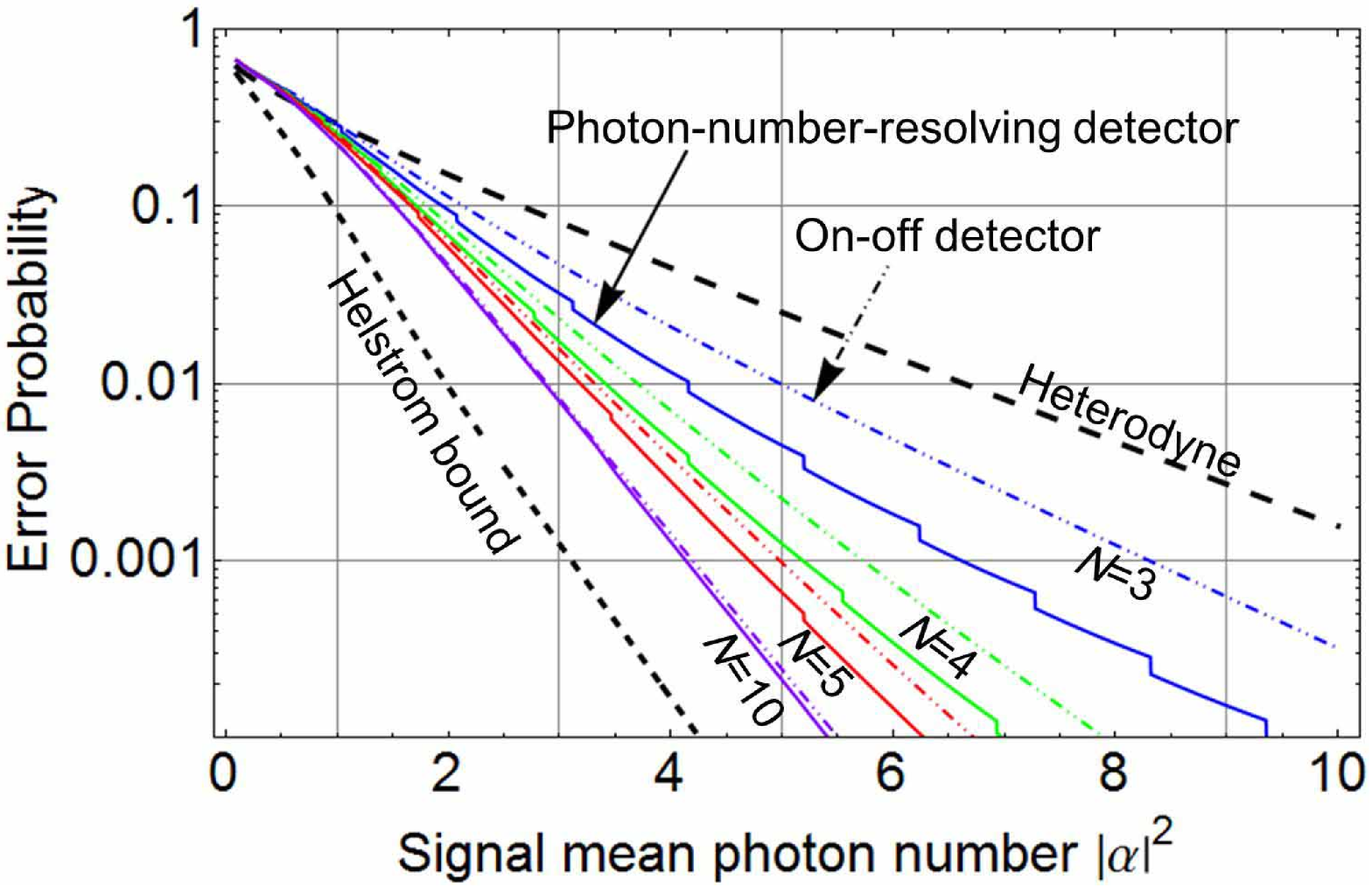}
\caption{(Color online) The displacement receiver consisting of $N$-step feedforward operations.
Solid and dot-dashed lines denote the error probabilities for the PNRD detection and the on-off detection respectively.
}
\label{DR_PNRD_DC_0_DE_1_n=10}
\end{center}
\end{figure}

Fig.~\ref{DR_PNRD_DC_0_DE_1_n=10} shows the error rates of the $N$-step feedforward receivers ($N=3,4,5,10$) with ideal PNRDs 
 ($\nu=0$, $\eta=1$, solid lines) and on-off detectors (dot-dashed lines).
Remarkably low error rates are obtained for the PNRD based receiver 
in the region of small $N$. 
For larger $N$, the performance of the both receivers almost coincide 
since multi-use of 
on-off detectors at feedforward effectively resolves 
the number of photons in the original signal. 
The error rates for the PNRD receiver show step-like curves. 
At each feedforward step, with a given outcome $n$, 
we chose $\alpha_m$ which maximizes $\Gamma(\alpha_{m}|n)$ 
as the next nulling signal. 
In other words, the feedforward behavior highly depends on 
the classification $\{ n | \Gamma(\alpha_m|n) \ge \Gamma(\alpha_l|n) \, 
(l \ne m) \}$. 
Due to the discrete nature of photon number, such a classification 
varies discretely as a function of $|\alpha|^2$ resulting in 
the step-like curves on the averaged error performance.

\begin{figure}[b]
\centering
\subfigure
{
\includegraphics[width=0.95\linewidth]
{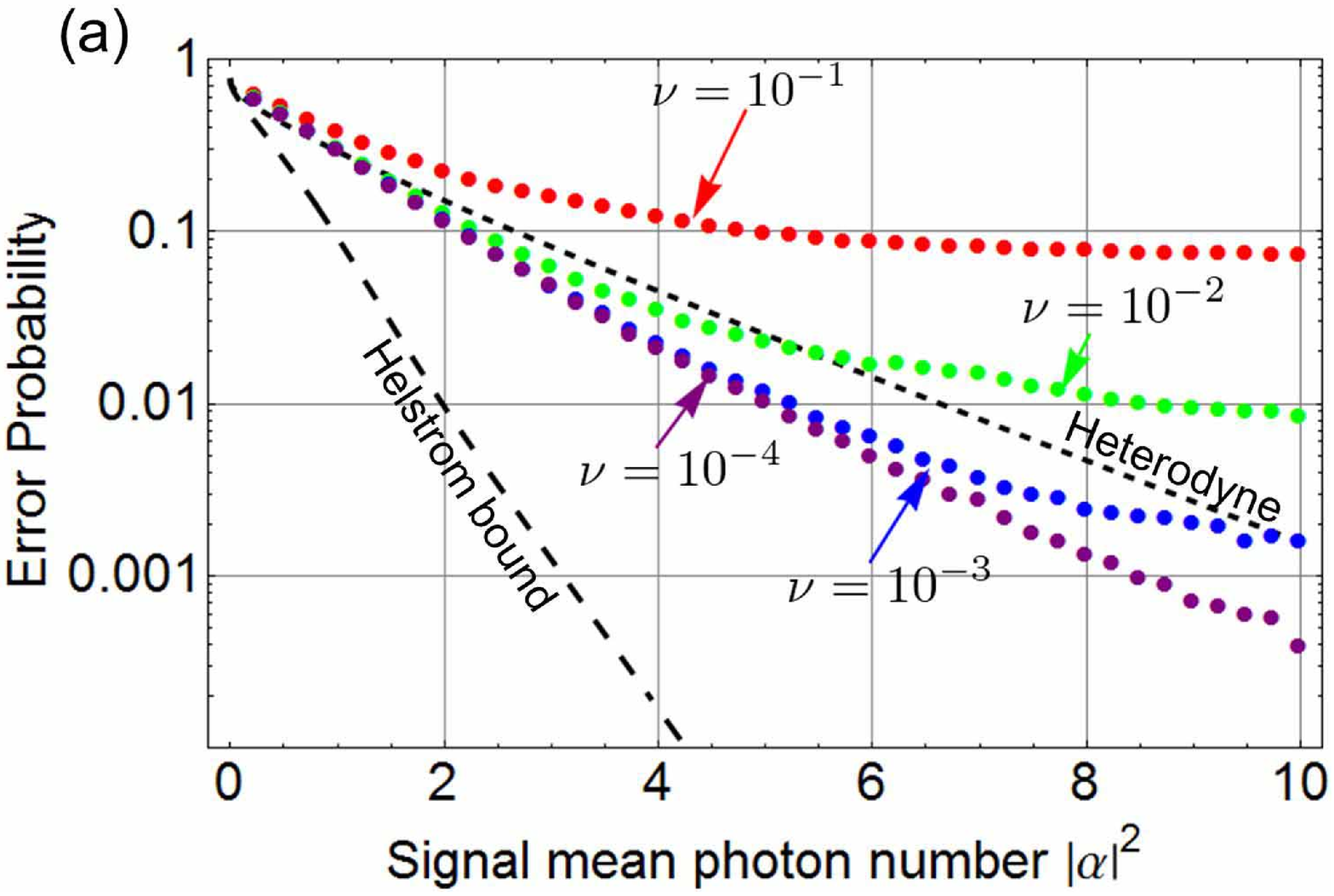}
\label{montecarlo_on_off_DC_vari_DE_1_0.25_step}
}
\subfigure
{
\includegraphics[width=0.95\linewidth]
{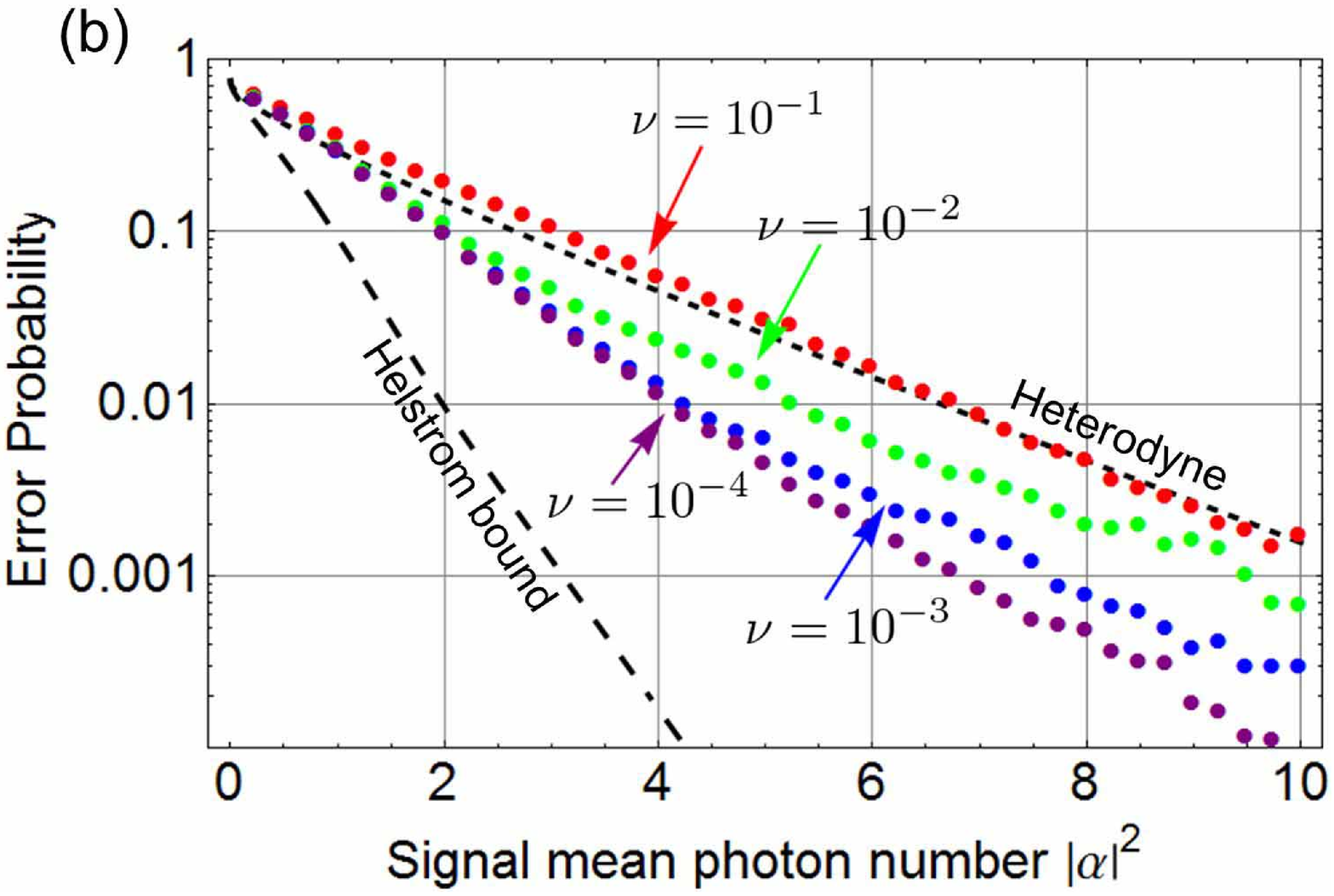}
\label{montecarlo_PNRD_DC_vari_DE_1_0.25_step}
}
\caption{(Color online) 
Degradation of the error rates depending on the dark count probability $\nu$ for 
(a)\,on-off detector, and
(b)\,PNRD.
The feedforward steps and the detection efficiency are fixed at $N=3$ and $\eta=1$ in both figures.
Each plot is given by a Monte Carlo simulation with $10^5$ trials. 
Non-monotonic fluctuations of the plots at relatively high $|\alpha|^2$ 
are due to the statistical errors of the simulation. 
}
\label{DC_vari_DE_1}
\end{figure}

For non-zero $\nu$, on the other hand, the analytical derivation of 
the error rate is almost intractable since all 
$\Gamma(\alpha_{m_{j}}|n_{j})$ could remain finite even after the $j$th stage. 
We therefore evaluate the error performance with non-ideal detectors by 
Monte Carlo simulations. 
Fig.~\ref{DC_vari_DE_1}(a) and (b) show the error rates for the on-off detectors and the PNRDs, respectively. 
We examined the performance for various $\nu$ with $\eta=1$ and $N=3$. For the on-off detectors 
(Fig.~\ref{montecarlo_on_off_DC_vari_DE_1_0.25_step}), 
the error rates are saturated at $P_e \approx \nu$, which implies that 
the dark counts seriously limits the performance of the receiver. 
On the other hand, the PNRD based receiver is clearly free from the
saturation problem(Fig.~\ref{montecarlo_PNRD_DC_vari_DE_1_0.25_step}). Since PNRD can discriminate incident photon numbers,
it can exclude the events for dark counts from those for real signals to
a certain extent, especially in the region where $\abs{\alpha}^2 \gg \nu$.
This reflects the robustness of the PNRD based receiver against the dark counts.
Apparently, this is impossible by on-off detectors under condition having 
the same number of feedforward steps.
Note that the robustness 
against the dark counts could be observed even with the on-off detectors 
if one allows large $N$ since it effectively provides the number resolving 
ability as mentioned above. 

\begin{figure}[h]
\centering
\subfigure
{
\includegraphics[width=0.95\linewidth]
{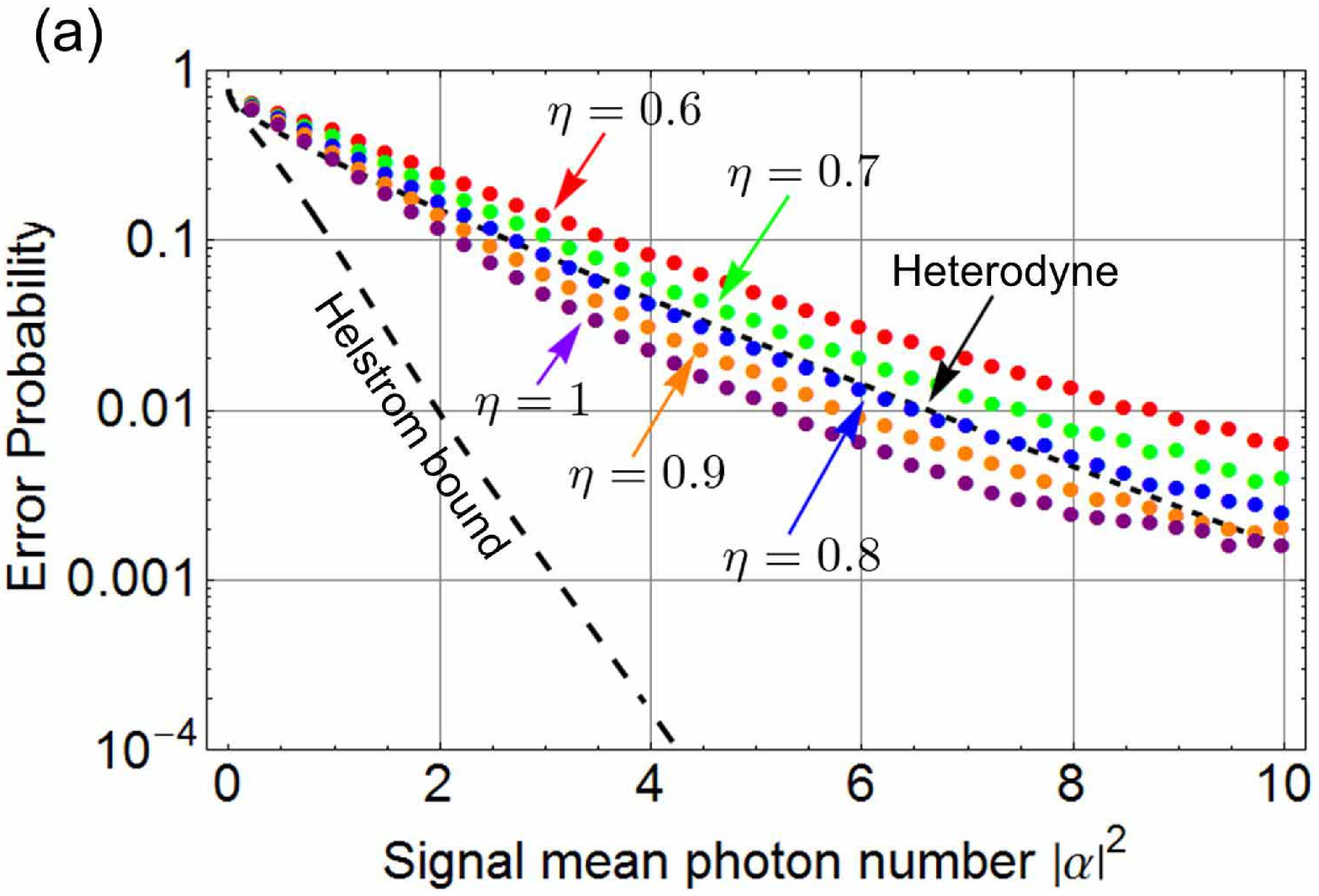}
\label{montecarlo_on_off_DC_10^(-3)_DE_vari_0.25_step}
}
\subfigure
{
\includegraphics[width=0.95\linewidth]
{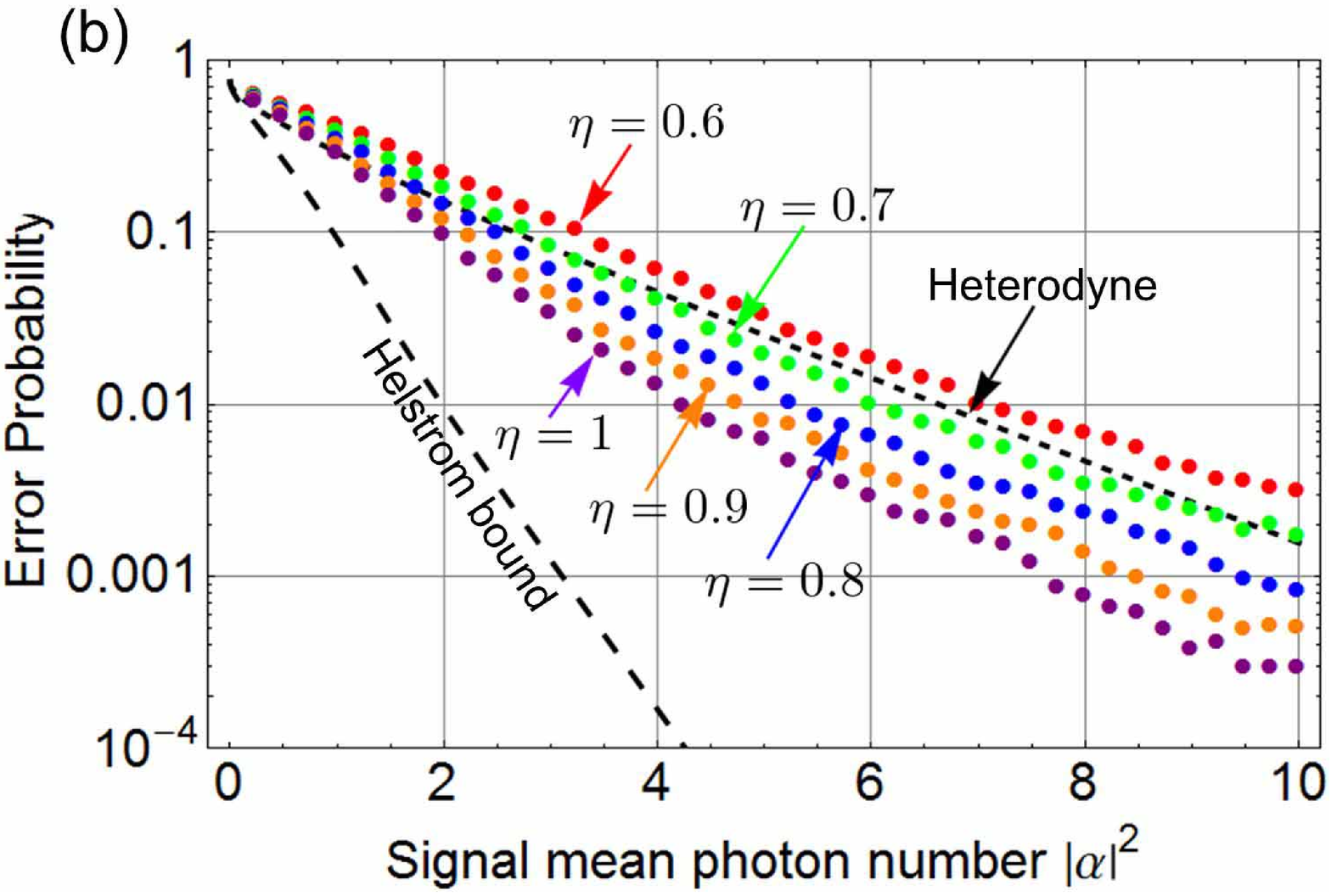}
\label{montecarlo_PNRD_DC_10^(-3)_DE_vari_0.25_step}
}
\caption{(Color online) 
Degradation of the error probabilities depending on the detection efficiency $\eta$ for 
(a)\,on-off detector,
(b)\,PNRD.
The feedforward steps and the dark count probability are fixed at $N=3$ and $\nu=10^{-3}$.}
\label{DC_10^(-3)_DE_vari}
\end{figure}

We also evaluate the dependence on the detector efficiency $\eta$ with $\nu=10^{-3}$ and $N=3$. 
Fig.~\ref{montecarlo_on_off_DC_10^(-3)_DE_vari_0.25_step} shows that the detector efficiency $90\%$ is at least required for the on-off detector to obtain the performance beyond the heterodyne limit, however, the requirement for the detector efficiency can be decreased to $70\%$ by employing the PNRDs in place of the on-off detectors.

\section{Conclusions}\label{Sect:4}

We proposed two quantum receivers for $M$-ary coherent state discrimination. 
The first one consists of displacements, on-off detectors, 
and squeezers without feedforward (or feedback) operations. 
The theoretical results showed that squeezing operation can slightly increase 
the performance at the weak signal region compared to the scheme without feedforward 
nor squeezing, presented in \cite{izumi2012}. 

The second one consists of displacements, PNRDs, and feedforward operations. 
We numerically demonstrated that, for the fixed number of feedforward steps, 
the PNRD based receiver shows sufficiently better 
performance than the on-off detector based receivers \cite{Becerra_NIST_2011_MPSK_emulation_experiment,izumi2012,Becerra13}. 
In other words, PNRD can decrease the number of feedforward steps. 
It was also shown that the PNRD receiver is robust against the dark count which strictly limits the performance of 
photon counting based receivers in a relatively higher photon number regime. 
Though our analyses are concentrated on the QPSK signals, 
we emphasize that we can generalize these receivers to the $M$-ary 
signals ($M>4$) in a straightforward way. 

Our results show that the PNRD receiver is a feasible scheme 
with the current technology and could achieve smaller error rates 
with a reduced number of feedforward steps. 
Fewer feedforward steps will allow us to detect the 
shorter pulsewidth or higher repetition rate signals, which are an important 
figure of merit toward implementing the practical application of quantum receivers. 

Finally, from a theoretical point of view, an interesting future issue 
is to investigate how to fill the gap between the ideal performance of 
the feedforward based receiver and the exact Helstrom bound for 
the QPSK signals. It could be achieved by the additional nonlinear process 
e.g. replacing the beamsplitters with higher order nonliner couplings.

\section{Acknowledgement}\label{Sect:5}
This work was supported by the Founding Program for World-Leading Innovative R\&D on Science and Technology (FIRST).


\appendix
\section{Displacement receiver with squeezing operation: 
Formulation}


In this appendix, we describe detailed derivations of 
$|\Psi_i\rangle$ and $\Pi_i$ discussed in Sec.~II, 
which are necessary to calculate Eq.~(\ref{eq:success_prob}). 
As illustrated in Fig.~\ref{Scheme_4PSK_3port}, 
the signal is split into three ports and at each port, 
displaced and squeezed before the on-off detection. 
Thus the state just before the detection is generally in 
the squeezed coherent state $\ket{\xi ; \beta}$ .
It is described in photon number bases as,
\beqa
\ket{\xi ; \beta}&=&\hat{S}(\xi)\hat{D}(\beta)\ket{0}
\nonumber
\\
&=&
\mathrm{e}^{-\frac{\abs{\beta^2}}{2}+\beta^2 \frac{\kappa^{\ast}}{2\mu}}\sum_{n=0}^{\infty}\frac{1}{\sqrt{n!\mu}}\Biggl[\frac{\kappa}{2\mu}\Biggr]^{\frac{n}{2}} H_n (\frac{\beta}{\sqrt{2\mu\kappa}})\ket{n},
\nonumber
\\
 \label{eq:2}
\eeqa
where $H_n$ is the $n$th Hermite polynomial, $\mu=\cosh{r}$ and $\kappa=\mathrm{e}^{i\phi}\sinh{r}$ \cite{MandelWolf}. 
Let $\beta_j$ and $\xi_j$ be the displacement and squeezing 
parameters, respectively, at port $j=A, B, C$. 
After applying the displacements and squeezings, $|\alpha_m\rangle$ 
is transformed to a three-mode state $|\Psi_m\rangle$ where 
\begin{eqnarray}
\ket{\Psi_0}_{ABC}
&=&\ket{\xi_{A};\beta_{A}}_A 
\nonumber
\\
&\otimes & \ket{\xi_{B};\sqrt {(1-R_1)R_2}(\alpha_0 -\alpha_2 )+\beta_{B}}_B
\nonumber
\\
&\otimes & \ket{\xi_{C};\sqrt {(1-R_1)(1-R_2)}(\alpha_0-\alpha_1 )+\beta_{C}}_C,
\nonumber
\\
\label{phi0}
\end{eqnarray}

\begin{eqnarray}
\ket{\Psi_1}_{ABC}
&=&\ket{\xi_{A};\sqrt{R_1}(\alpha_1-\alpha_0 )+\beta_{A}}_A
\nonumber
\\
&\otimes & \ket{\xi_{B};\sqrt {(1-R_1)R_2}(\alpha_1 - \alpha_2)+\beta_{B}}_B 
\nonumber
\\
&\otimes & \ket{\xi_{C};\beta_{C}}_C,
\label{phi1}
\\
\ket{\Psi_2}_{ABC}
&=&\ket{\xi_{A};\sqrt{R_1}(\alpha_2- \alpha_0)+\beta_{A}}_A
\otimes \ket{\xi_{B};\beta_{B}}_B
\nonumber
\\
&\otimes & \ket{\xi_{C};\sqrt {(1-R_1)(1-R_2)}(\alpha_2 - \alpha_1 )+\beta_{C}}_C,
\nonumber
\\
\label{phi2}
\\
\ket{\Psi_3}_{ABC}
&=&\ket{\xi_{A};\sqrt{R_1}(\alpha_3 - \alpha_0 ) +\beta_{A}}_A
\nonumber
\\
&\otimes & \ket{\xi_{B};\sqrt {(1-R_1)R_2}(\alpha_3 - \alpha_2 )+\beta_{B}}_B 
\nonumber
\\
&\otimes & \ket{\xi_{C};\sqrt {(1-R_1)(1-R_2)}(\alpha_3 - \alpha_1 )+\beta_{C}}_C.
\nonumber 
\\
\label{phi3}
\end{eqnarray}
Note that the parameters optimized in Sec.~II are 
$R_1$, $R_2$, $\beta_A$-$\beta_C$, and $\xi_A$-$\xi_C$.

An on-off detector only discriminates 
zero or non-zero photons. 
Its measurement operators are given by 
\beqa\label{Imperfect_on_off_detector}
\hat{\Pi}_\mathrm{off}
&=&e^{-\nu}\displaystyle\sum^{\infty}_{n=0}(1-\eta)^n\ket{n}\bra{n}\;,
\label{eq:3}\\
\hat{\Pi}_\mathrm{on}&=&\hat{I}-\hat{\Pi}_\mathrm{off}\;,
\label{eq:4}
\eeqa
where $\nu$ is the dark count probability 
and $\eta$ is the detection efficiency.
In our scheme three on-off detectors are used and 
the signal decision is carried out by 
the following combination of detection outcomes:
\begin{equation}
\begin{array}{llll}
\hat \Pi_0
=\hat \Pi_\mathrm{off}^A \otimes \hat I^B \otimes \hat I^C\;,
\\
\hat \Pi_1
=\hat \Pi_\mathrm{on}^A \otimes \hat \Pi_\mathrm{off}^B \otimes \hat I^C\;,
\\
\hat \Pi_2
=\hat \Pi_\mathrm{on}^A \otimes \hat \Pi_\mathrm{on}^B \otimes \hat \Pi_\mathrm{off}^C\;,
\\
\hat \Pi_3
=\hat \Pi_\mathrm{on}^A \otimes \hat \Pi_\mathrm{on}^B \otimes \hat \Pi_\mathrm{on}^C\;,
\end{array}
\label{eq:6}
\end{equation}
where $\hat{I}$ is an identity operator. 
These descriptions allow us to calculate the probability 
$\langle\Psi_i|\Pi_j|\Psi_i\rangle$. 
In general, 
the probability of having an ``off'' outcome for the squeezed coherent state $\ket{\xi ; \alpha_{m}}$ is given by 
\beqa
P_{\mathrm{off}}
&=&
\bra{\xi ; \alpha_{m}}\hat{\Pi}_\mathrm{off}\ket{\xi ; \alpha_{m}}
\nonumber
\\
&=&
\mathrm{e}^{-\nu-\alpha^2 \left\{ 1-\tanh{r} \cos{(\frac{2m+1}{2}\pi-\phi ) }\right\}}
\nonumber
\\
&\times &
\sum_{n=0}^{\infty} \frac{(1-\eta)^n}{n! \mu}\Biggl[\frac{\abs{\kappa}}{2\mu}\Biggr]^n \abs{H_n (\frac{\alpha}{\sqrt{2\mu\kappa}})}^2 .
\label{eq:5}
\eeqa


\end{document}